\documentclass[preprint,aps,prb,showpacs,amsfonts,amssymb,superscriptaddress]{revtex4}
\usepackage{graphicx}

\begin{document}
\title{Intersubband electroluminescent devices operating in the strong coupling regime}
\author{P. Jouy}
\affiliation{Laboratoire ``Mat\'eriaux et Ph\'enom\`enes
Quantiques'', Universit\'e Paris Diderot-Paris 7, CNRS-UMR 7162, 75013
Paris, France}
\author{A. Vasanelli}
\affiliation{Laboratoire ``Mat\'eriaux et Ph\'enom\`enes
Quantiques'', Universit\'e Paris Diderot-Paris 7, CNRS-UMR 7162, 75013
Paris, France}
\author{Y. Todorov}
\affiliation{Laboratoire ``Mat\'eriaux et Ph\'enom\`enes
Quantiques'', Universit\'e Paris Diderot-Paris 7, CNRS-UMR 7162, 75013
Paris, France}
\author{L. Sapienza}
\affiliation{Laboratoire ``Mat\'eriaux et Ph\'enom\`enes
Quantiques'', Universit\'e Paris Diderot-Paris 7, CNRS-UMR 7162, 75013
Paris, France}
\author{R. Colombelli}
\affiliation{Institut d'Electronique Fondamentale, Universit\'e
Paris Sud, CNRS-UMR 8622, 91405 Orsay, France}
\author{U. Gennser}
\affiliation{Laboratoire de Photonique et Nanostructures,
LPN-CNRS, Route de Nozay, 91460 Marcoussis, France}
\author{C. Sirtori}
\affiliation{Laboratoire ``Mat\'eriaux et Ph\'enom\`enes
Quantiques'', Universit\'e Paris Diderot-Paris 7, CNRS-UMR 7162, 75013
Paris, France}


\begin{abstract}
We present a detailed study of the electroluminescence of intersubband devices operating in the light-matter strong coupling regime. The devices have been characterized by performing angle resolved spectroscopy that shows two distinct light intensity spots in the momentum-energy phase diagram. These two features of the electroluminescence spectra are associated with photons emitted from the lower polariton branch and from the weak coupling of the intersubband transition with an excited cavity mode. The same electroluminescent active region has been processed into devices with and without the optical microcavity to illustrate the difference between a device operating in the strong and weak coupling regime. The spectra are very well simulated as the product of the polariton optical density of states, and a function describing the energy window in which the polariton states are populated. The voltage evolution of the spectra shows that the strong coupling regime allows the observation of the electroluminescence at energies otherwise inaccessible.
\end{abstract}
\pacs{42.50.-p, 73.63.Hs, 78.60.Fi, 85.60.Jb}

\maketitle
\section{Introduction}
Intersubband transitions in semiconductor quantum wells are the mechanism at the heart of unipolar devices, such as quantum well infrared photodetectors and quantum cascade lasers. As the energy difference between two subbands mainly depends on the thickness of the quantum well, such devices can be designed to operate in a broad frequency range. As an example, quantum cascade lasers presently cover a large wavelength span  between 2.6~$\mu$m and 250~$\mu$m.~\cite{cathabard,walther} In these devices population inversion, and thus the optical gain, is obtained thanks to bandstructure engineering, by tailoring the subband lifetimes. On the contrary, the realization of light emitting devices in this frequency range is limited by the poor value of the radiative quantum efficiency. Typical values of the spontaneous emission lifetime are of the order of 10 - 100 ns, while the nonradiative lifetime is of the order of ps. The implementation of the light-matter strong coupling concepts within intersubband electroluminescent devices has been  seen recently as a promising way to increase the radiative quantum efficiency with respect to devices operating in the usual weak coupling regime~\cite{colombelli, deliberato_PRB2008}.
The first demonstration of the strong coupling regime between an intersubband excitation and a microcavity photonic mode was obtained in 2003 by reflectivity measurements~\cite{dini}. The quasi particles issued from this coupling are called intersubband polaritons. The strength of the coupling is in this case an important fraction of the photon energy~\cite{ciuti_PRB2005}; it depends on the electronic density in the fundamental subband~\cite{anappara_APL2005} and on the electron effective mass~\cite{anappara_ssc}. Furthermore, in this kind of systems, an unprecedented ultra-strong coupling regime can be attained in the mid-infrared~\cite{anappara_ultrastrong} and in the THz frequency range~\cite{todorov_PRL}. The possibility of merging the subband engineering typical of quantum cascade lasers and the properties of intersubband polaritons has been exploited soon after their first observation, by the realization of mid-infrared photodetectors operating in the strong coupling regime~\cite{dupont, sapienza_APL}.

A mid-infrared light emitting device based on intersubband polaritons and working up to room temperature has been recently demonstrated~\cite{sapienza_PRL}. In this device, the subband engineering led to a selective electronic injection into the polariton states~\cite{todorov_APL}, allowing a frequency tunability of the electroluminescence of $\approx 20\%$. On the theoretical side, several efforts have been made to describe electroluminescence from polaritonic devices. In fact, the interplay between fermionic transport and bosonic polaritons makes the system quite complex to model. In ref.~\onlinecite{ciuti_PRA} this difficulty is bypassed by considering, instead of an electronic injector, the coupling between the polaritons and a dissipation bath of electronic excitations. This allows the authors to obtain an analytical expression for the electroluminescence. Electroluminescence from intersubband polaritons has also been described within a completely fermionic approach, in the case either of a broad band~\cite{deliberato_PRB2008}, or of a narrow band~\cite{deliberato_PRB2009} injector. The  electroluminescence spectra calculated using this second approach are similar to those of refs.~\onlinecite{sapienza_PRL, todorov_APL}.

In this work we present a detailed study of the electroluminescence from an intersubband device working in the light-matter strong coupling regime. In order to reveal the peculiar features coming from the polariton dispersion in the electroluminescence spectra, they are compared to those obtained from an identical device, but with a different photonic confinement that hinders the operation in the strong coupling regime. We show that the electroluminescence signal from the polaritonic device is composed by two main contributions: the first one comes from the lower polariton branch; the second one is due to the weak coupling of the intersubband transition with an excited cavity mode. The electroluminescence spectra are calculated by considering that polariton states are only populated within an energy window associated to the electronic injector. By changing the bias applied to the device, we show that the strong coupling regime allows the enhancement of the electroluminescence signal at an energy that depends on the electronic injector.

The paper is organized as follows. In section~\ref{sample} we present the two samples studied in this work. The simulated absorption spectra for both samples are discussed and demonstrate that by changing the cavity the same electroluminescence active region can move from the strong to the weak coupling light-matter regime. In section~\ref{spectra} we present the electroluminescence spectra measured for both samples. Section~\ref{simul} is devoted to the simulation of the electroluminescence spectra, based on our model~\cite{sapienza_PRL}. In section~\ref{voltage} we present the voltage evolution of the electroluminescence spectra and their simulation. Finally, in section~\ref{enhancement} we present the experimental and simulated quantum efficiency and a discussion on the enhancement of the spontaneous emission in our device. Conclusions are drawn in section~\ref{concl}.

\section{Samples}
\label{sample}
The active region of our devices consists of a GaAs/Al$_{0.45}$Ga$_{0.55}$As quantum cascade structure, composed of 30 identical periods. Figure~\ref{bd} shows one period of the cascade, simulated at a voltage of 6~V by solving coupled Schr\"{o}dinger and Poisson equations. Each period contains a main quantum well, with two subbands (labeled 1 and 2) separated by $E_{21}=161$~meV, and an injection/extraction region. In the absence of the coupling with the cavity mode, electrons are electrically injected into the excited subband, and then relax radiatively or non-radiatively into the fundamental subband. Tunneling out from the quantum well into a miniband allows electron extraction and re-injection into the following period of the cascade. The structure design is such to increase the tunneling time out of the fundamental subband and avoid population inversion~\cite{colombelli}. The quantum cascade structure is grown onto a low-refractive index bi-layer, composed of a 0.56-$\mu$m-thick GaAs layer $n$-doped to $3 \times 10^{18}$ cm$^{-3}$, and a 0.52-$\mu$m-thick Al$_{0.95}$Ga$_{0.05}$As layer. The growth is terminated by a couple of $n$-doped GaAs layers, 86-nm and 17-nm-thick, respectively doped to $1 \times 10^{17}$ cm$^{-3}$ and $3 \times 10^{18}$.
In order to perform electroluminescence spectra, the sample is etched into circular mesas with a diameter of 200 $\mu$m. We fabricated two devices with the same active region but with different resonators. In the first device ({\textit{"polaritonic device"}}) light is confined between the low refractive index layers and a metallic mirror ([Ni(10nm)/Ge(60nm)/Au(120nm)/Ni(20nm)/Au(200nm)]) evaporated on the top of the mesa. In this device, the metallic mirror of the cavity is also the top contact. For the second device ({\textit{"weak coupling device"}}), the top contact is only 50 $\mu$m diameter (see inset of fig.~\ref{topEL}  for a top view of the mesa devices). In this way only 6$\%$ of the mesa surface is covered with gold and barely contribute to the optical confinement. However, the current injection into the quantum cascade structure is still possible, thanks also to the heavily doped top layer that allows lateral current spreading.

The dispersion of the photonic mode of the polaritonic device, obtained by using transfer matrix formalism, is shown in figure~\ref{absorption}a. Here the absorption coefficient of the cavity is plotted in color scale, as a function of the photon energy and of the in-plane momentum ($k_{//}$). The absorption coefficient has been calculated as $1-R$, where $R$ is the reflectivity, since in our experiment there is no transmission through the upper mirror~\cite{sapienza_PRL, ciuti_PRA}. The resonance condition, allowing for a strong coupling between the intersubband transition and the cavity mode, is fulfilled for a value of the in-plane photon momentum of approximately $k_{res}=2.55 \, \rm{\mu m}^{-1}$. It is related to the photon energy $E_p$ and to the internal angle for light propagation $\theta$ by the following formula \cite{sapienza_PRL}:
\begin{equation}
\label{kpar}
k_{//}=E_p \, \frac{n_s}{\hbar c} \sin \theta
\end{equation}
where $n_s$ is the substrate refractive index. By replacing the values of the intersubband transition energy $E_{21}$ and that of $k_{//}$ at resonance, we obtain the corresponding value of the internal angle for light propagation, $\theta_{res}=72.5^{\circ}$. In order to observe the light-matter coupling in angle resolved experiments, we polished the facet of the sample at $70^{\circ}$. In this way the light propagating angle inside the cavity can be varied between $58^{\circ}$ and $83.5^{\circ}$, hence spanning the interesting angular range for the observation of the strong coupling regime.
In figure~\ref{absorption}a, one can also notice a second order cavity mode, which is much broader than the fundamental one. The two modes are clearly resolved in the inset of fig.~\ref{absorption}a, where the simulated absorption (in logarithmic scale) at $E_p=160$~meV is plotted as a function of $k_{//}$. The full width at half the maximum (FWHM) of the second order mode is $\approx 25$ meV at the intersubband transition energy, as opposed to $\approx 3$ meV for the fundamental mode. The second order mode is resonant with the intersubband transition for a value of the in-plane photon momentum $k_{exc} \approx 2 \, \rm{\mu m}^{-1}$. It follows that, in an angle resolved measurement, its contribution can be observed for angles close to $55^{\circ}$.
Figure~\ref{absorption}b shows the calculated absorption spectrum for the (unbiased) device in the same angular range spanned in the experiments. The contribution of the intersubband transition has been taken into account in the dielectric permittivity of the quantum-well layers including an additional term in the form of an ensemble of classical polarized Lorentz oscillators \cite{dini}. The dispersions of the Au \cite{ordal} and of the doped layers \cite{palik} have also been included. In the simulations, we used $E_{21} =161$ meV for the energy of the bare intersubband transition and $N_1 =6 \times 10^{11}\, \rm{cm}^{-2}$ for the electronic density in the fundamental subband. This value gives the best agreement between the simulated and the measured angle resolved absorption spectra \cite{sapienza_PRL} and determines the value of the vacuum Rabi splitting in the system: $2\hbar \Omega_R=11$ meV. Figure~\ref{absorption}b shows a clear anticrossing between the intersubband excitation and the cavity mode, giving rise to the two polariton branches. For $k_{//} \approx k_{exc}$ we observe a second feature in the calculated absorption spectrum at the energy of the bare intersubband transition, due to the coupling between the intersubband transition and the excited cavity mode. This coupling is weak due to the broadening of the cavity mode. In fact, the condition for the system to enter the light-matter strong coupling regime is ${\Omega_R}^2> \Gamma_{12} \Gamma_{cav}$ , where $\Gamma_{12}$ is the non radiative broadening of the intersubband transition (in our case $\Gamma_{12}=9$ meV, as extracted from the bare electroluminescence spectrum) and $\Gamma_{cav}$ is the cavity mode broadening.

Figure~\ref{absorption_nomirror}a shows the dispersion of the cavity mode in the weak coupling device, obtained using exactly the same parameters as in fig.~\ref{absorption}a. We can see that there is still a cavity effect, between the air and the low refractive index layers. The simulated absorption spectrum, including the contribution of the intersubband transition, is shown in fig.~\ref{absorption_nomirror}b. In this case the intersubband transition is only weakly coupled to the photonic mode.

\section{Electroluminescence spectra in weak and strong coupling}
\label{spectra}
After fabrication and mechanical polishing of the facet to the proper angle, the sample is indium soldered onto a copper holder and mounted in a cryostat for angle resolved electroluminescence measurements. All the experimental results in this work have been obtained at 77 K. The electroluminescence signal from the substrate is collected with a $f/2$ ZnSe lens, analyzed by a Fourier Transform Infrared Spectrometer and detected using a HgCdTe detector through a $f/0.5$ ZnSe lens. The angular resolution of the optical system has been estimated to be approximately $0.5^{\circ}$.
Figure~\ref{topEL} presents the electroluminescence spectra of the device with (left panel) and without (right panel) the top Au layer, measured at the same current (14 mA) and voltage (4.5 V) and for two different values of the internal angle of light propagation. The spectra from the weak coupling device are very similar at both angles. On the contrary, in the left panel of figure~\ref{topEL} we show that the EL spectrum from the polaritonic device measured at $71.8^{\circ}$, i.e. close to $\theta_{res}$, is very different from the one obtained far from resonance (52.4$^\circ$). The spectra from the weak coupling device consist of a main peak, centered at the energy $E_{21}$, and a low energy tail. This shape is typical of the electroluminescence spectrum of a quantum cascade structure below the alignment voltage~\cite{benveniste}. It can be interpreted as the sum of two contributions. The main peak is due to the radiative transition from the excited to the fundamental subband (displaying in this case a Gaussian lineshape), while the low energy tail comes from the diagonal radiative transition of the injector state (labelled {\textit{inj}} in the band diagram in fig.~\ref{bd}) to the fundamental subband, at an energy $E_{inj}$. This contribution can be fitted with a Gaussian distribution, whose energy position and width $\sigma$ depend on the applied voltage.

The EL spectra of both structures measured at an internal angle close to $55^{\circ}$ are quasi identical. This is consistent with the fact that for these angles, the polaritonic structure is very far from its resonant condition. The light observed comes from the previously mentioned high order cavity mode, weakly coupled with the intersubband transition. As a consequence, the injector energy can also be inferred by analyzing the EL spectra from the polaritonic device at low angle. This method has the advantage to guarantee identical voltage and current conditions for the polaritonic and for the weakly coupled spectra.

Figure~\ref{fits} shows electroluminescence spectra (symbols) from the polaritonic device, measured at different voltages at an angle of $58^\circ$. They are very well fitted (continuous line) by using a sum of two Gaussian functions (dashed lines)~\cite{nota_fit}. The main peak is centered at the same energy in the entire voltage range ($E_{21}=161$~meV), while the energy position of the injector increases with the voltage and approaches $E_{21}$. The values of $E_{inj}$ and $\sigma$ obtained from the fit are summarized in table~\ref{tabella} (second and third columns). We can see that the width of the Gaussian function describing the injector monotonically decreases. This behavior is typical of a tunnel coupling between the injector and the excited state of the radiative transition in a quantum cascade structure~\cite{benveniste}.

\begin{table}
\begin{center}
\begin{tabular}{|c|c|c|c|c|}
\hline
Voltage (V) & $E_{inj}^{fit}$ (meV) & $\sigma^{fit}$ (meV) & $E_{inj}$ (meV) & $\sigma$ (meV)\\
\hline
4.5 & 151.6 & 10 & 150.5 & 12\\
5 & 152.9 & 8.5 & 154 & 9\\
6 & 156.2 & 7.2 & 157 & 7\\
7.75 & 158 & 6.4 & 158 & 6.5\\
13 &  160 & 5.2 & 160 & 6\\
\hline
\end{tabular}
\caption{Energy position and width of the diagonal transition from the injector to the fundamental subband  of the main quantum well, as extracted from the fit of the electroluminescence spectra measured at $\approx 58^{\circ}$ internal angle (second and third column). The fourth and fifth column presents the values of $E_{inj}$ and $\sigma$ used to simulate the polaritonic electroluminescence (see section~\ref{voltage}). } \label{tabella}
\end{center}
\end{table}

We display contour plots of the electroluminescence as a function of the photon energy and in-plane momentum from $\approx 40$ angle resolved electroluminescence spectra, and through the use of eq.~\ref{kpar}. The comparison between the contour plots obtained for the two devices at a voltage of 4.5~V (14 mA) is shown in fig.~\ref{EL_comp}. In the polaritonic device (fig.~\ref{EL_comp}a), the most important contribution to the electroluminescence signal comes from the lower polariton branch~\cite{todorov_APL}. This contribution is associated to a tunneling of the electrons from the injector state directly into the polariton states~\cite{sapienza_PRL}. On the contrary, for the weak coupling device (fig.~\ref{EL_comp}b), the electroluminescence signal is centered at the energy of the intersubband transition. Note that an electroluminescence signal at the bare intersubband transition energy is observed, even in the polaritonic device, because of the coupling with the high order cavity mode. This is analogous to what observed in reflectivity measurements by Dupont {\textit{et al.}}~\cite{dupont_2007}: in their work the coupling between the intersubband transition and the excited cavity mode is responsible for an intense absorption peak between the upper and lower polariton branches. This effect has been theoretically investigated by Za{\l}u\.{z}ny {\textit{et al.}~\cite{zaluzny1, zaluzny2}

\section{Simulation of the electroluminescence from the polariton states}
\label{simul}
The theoretical description of the electroluminescence from intersubband polaritons is a complex problem, because the subbands are coupled not only to the radiation, but also to the injection and extraction minibands of the quantum cascade structure. This problem has been addressed recently in several publications (see for example ref.~\onlinecite{deliberato_PRB2009} and references therein), following different approaches. The first attempt to calculate the electroluminescence spectrum from an intersubband polariton device has been proposed in ref.~\onlinecite{ciuti_PRA}. In this article, the authors describe the input-output dynamics of an optical cavity in the light-matter ultra-strong coupling regime. They consider the case of an intersubband excitation in a cavity coupled to two bosonic dissipative baths, for the photon field and for the electronic excitation. Within this model, the authors derive an analytical expression for the electroluminescence spectrum, which is proportional to a term accounting for the spectral shape of the electronic reservoir. Based on this result, a phenomenological model was used in ref.~\onlinecite{sapienza_PRL}, in which the experimental electroluminescence spectrum was simulated as the product of the absorption spectrum (in the strong coupling regime) and of a Gaussian function describing the injector state. In this case the electronic excitation bath may be considered as pairs of electrons tunneling in and out the main quantum well. The physical meaning of this model is that the absorption spectrum describes the optical density of states of the polaritonic system, while the Gaussian function selects the energy window in which the polariton states are populated. This description is analogue to that of the photoluminescence from microcavity exciton polaritons. Indeed, it is described as the occupancy of the polariton states times the coupling out of such polaritons, i.e. the reverse process of incoupling of outside photons, given by the absorption coefficient~\cite{stanley, weisbuch_review2000, houdre_PRL, houdre_review}. Furthermore, the fact that in our system the injector acts as a filter for the polaritonic emission has also recently been  obtained within a completely fermionic model, describing an electronic injection into an excited subband in a system strongly coupled with light \cite{deliberato_PRB2009}.

Following ref. \onlinecite{sapienza_PRL}, the electroluminescence spectrum is proportional to:
\begin{equation}
\mathcal{L}(E)=A_{N_1}(E) \times \exp \left(-\frac{\left(E-E_{inj}\right)^2}{2\sigma^2}\right)
\end{equation}
$A(E)$ is the calculated absorption spectrum; the index $N_1$ is inserted to remind the dependence of the vacuum Rabi splitting (hence of the absorption spectrum) from the square root of the population density on the fundamental subband (the population on the upper subband is always negligible). $E_{inj}$  and $\sigma$ are inferred from the electroluminescence spectrum measured at an angle close to $55^{\circ}$ (see table~\ref{tabella}) and slightly adjusted to reproduce the experimental data. In order to compare the simulated and experimental electroluminescence, we have to take into account the effect of the reflection on the polished facet, by computing the Fresnel coefficients. In fact the intensity emitted from the facet $I_{out}$ is proportional to that incident to the facet $I_{in}$ times the Fresnel coefficient $T=\left( 4 n_s \cos \theta_{out} \cos \theta_{in} \right) / {\left( n_s \cos \theta_{out} + \cos \theta_{in} \right)}^2$, where $\theta_{in}$ and $\theta_{out}$ are respectively the incident and refracted angle, measured with respect to the normal to the facet. The emitted power per unit angle is proportional to~\cite{lukosz, benisty}:
\begin{equation}
\frac{d I_{out}}{d \theta_{out}} \propto \frac{\cos^2 \theta_{out}}{\left( n_s \cos \theta_{out} + \cos \theta_{in}\right)^2}
\end{equation}
with $\theta_{in}=\alpha - \theta$, $\alpha$ the polishing angle of the facet and $\theta$ the internal angle for light propagation. In order to obtain the optical power collected by the detector, we multiply the emitted intensity per unit angle by the apparent surface of the facet in the direction of observation (which is proportional to $\cos \left( \theta_{out}\right)$) and by the projection of the surface of the mesa on the polished facet (which is proportional to $\cos \left( \theta \right)$). Finally, the electroluminescence spectra are simulated as:
\begin{equation}
\label{sim_EL}
\mathcal{L}_{sim}(E)=\mathcal{L}(E) \times \frac{\left(1-n_s^2 \sin^2\left(\alpha - \theta \right)\right)^{3/2} \cos \theta}{\left(n_s \left(1-n_s^2 \sin^2\left(\alpha - \theta \right) \right)^{1/2}+\cos \left(\alpha-\theta\right)\right)^2}
\end{equation}

Figure \ref{twosimul}a shows the calculated electroluminescence spectrum for an applied voltage of 4.5 V. The ground state population density used in the simulation is $N_1=4 \times 10^{11} \, \rm{cm}^{-2}$. The energy position of the injector is 150.5 meV; the Gaussian width $\sigma$ is 12 meV. Analogously to what has been observed in the experimental spectrum (fig.~\ref{EL_comp}a), we can distinguish two features in the simulation. The first contribution corresponds to the lower polariton branch, and it is spectrally limited within the energy window defined by the injector width. The second contribution is much broader and it is centered at the energy of the bare intersubband transition, as already discussed before. The observation of this electroluminescence signal at the energy $E_{21}$, due to the second order cavity mode, indicates that not all the electrons are injected into the lower polariton branch. This is an important limiting factor for the quantum efficiency of the device.
A comparison between figures \ref{EL_comp}a and \ref{twosimul}a shows that there is a very good agreement between the results of our model and the experimental spectra. The influence of the electronic density on the electroluminescence spectrum will be discussed in the next section.

In Fig. \ref{twosimul}b we show the contour plot obtained by taking a measured spectrum far from resonance (the one at $52.4^{\circ}$ shown in Fig.~\ref{topEL}) multiplied by the absorption calculated in Fig.~\ref{absorption} and by the Fresnel term in eq.~\ref{sim_EL}.  By comparing figures~\ref{twosimul}a and \ref{twosimul}b it is apparent that the first one reproduces very well the data, while the second one shows a very different result. This stresses the fact that to reproduce the data it is essential to consider the energy of the electrons that are injected into the polariton states of our system. In other words the polariton emission spectra cannot be reproduced by multiplying an internal source by the photon density.

\section{Voltage dependence of the electroluminescence spectra}
\label{voltage}
By changing the bias applied to the quantum cascade structure, the energy of the injector state varies with respect to the ground state of the quantum well. This strongly affects the electroluminescence spectrum and has been recently exploited to tune the electroluminescence of a device based on a single quantum well~\cite{todorov_APL}. Figure~\ref{panel_meas} shows the contour plots of the electroluminescence measured at 77~K at different voltages, from 4~V to 13~V. While increasing the applied voltage, two effects are clearly visible in these spectra: the electroluminescence from the lower polariton branch shifts towards higher energies; moreover the intensity of the electroluminescence at the energy $E_{21}$ becomes preponderant with respect to that from the lower polariton branch. These two effects are a direct consequence of the fact that the injector energy depends on the voltage applied to the device. The last two columns of table~\ref{tabella} summarize the values of $E_{inj}$ and $\sigma$ used to simulate the electroluminescence spectra, by using eq.~\ref{sim_EL}, for the different voltages. These values are very close to those obtained from the fit of the far from resonance spectra~\cite{nota}. The simulated spectra are shown in fig.~\ref{panel_sim}: they show an excellent agreement with the measured spectra reported in fig.~\ref{panel_meas}.
The simulation reproduces the entire dynamics of the system: the energy shift of the lower polariton luminescence and the increase of the weakly coupled  $E_{21}$ transition.  The ground state population density $N_1$, which affects the Rabi energy, is kept constant and equal to $4 \times 10^{11} \, \rm{cm}^{-2}$. This is consistent with our design of the quantum cascade structure, optimized to preserve a long tunneling time out from the fundamental subband. In a rate equation model, the ground state population $N_1$ is given by~\cite{colombelli}:
\begin{equation}
\label{rate}
N_1 = \frac{J}{q} \tau_{out}+N_s \exp \left( -\frac{\Delta}{k_B T} \right)
\end{equation}
where $J$ is the current density, $q$ is the electron charge, $\tau_{out}$ is the tunneling time from subband 1 into the injector states, $N_s$ is the total doping concentration minus the density of electrons that participate in the transport, and $\Delta$ is the energy difference between state 1 and the injector quasi-Fermi energy, as indicated in figure~\ref{bd}. From eq.~\ref{rate}, for a current of 2 A (corresponding to a voltage of 13~V), we estimate $\tau_{out} \approx 4-5$~ps. This is compatible with a tunneling out of the fundamental subband assisted by interface roughness scattering, which in our system gives $\approx 6$ ps \cite{leuliet}.

 A direct proof that the population density of the ground state stays constant over the voltage span of our experiment can be obtained by studying the electroluminescence at a fixed energy. In fact, for constant energy the two contributions to the optical signal clearly appear at different $k_{//}$; furthermore they only depend on the optical density of states, hence on the population density $N_1$. Figure~\ref{coupes}a shows the simulated electroluminescence at the energy of the intersubband transition ($E_{21}=161$ meV) as a function of the in-plane photon momentum, calculated for $N_1=4 \times 10^{11} \, \rm{cm}^{-2}$ (dashed line) and $N_1=1 \times 10^{11} \, \rm{cm}^{-2}$ (continuous line). For the highest value of the electronic density, the intersubband excitation is in the strong coupling regime with the fundamental cavity mode and in the weak coupling regime with the excited one. As a consequence, the absorption is inhibited close to $k_{res}$ and exalted close to $k_{exc}$. On the contrary for the lower value of the electronic density the intersubband excitation is weakly coupled with both cavity modes. As can be observed from fig.~\ref{coupes}a, the ratio between the intensity of the two resonances varies of more than a factor of 4. It is therefore apparent that a variation of the strength of the coupling directly affects the relative intensity of the two peaks. Figure~\ref{coupes}b shows the normalized electroluminescence spectra measured in the entire voltage range at the energy $E_{21}$. Within the noise all the spectra are identical: this means that the coupling constant has not changed for different voltages. Moreover, it shows that, apart from a multiplication factor, all the curves are identical for all voltages. This proves that the evolution of the electroluminescence spectra is entirely due to the position and shape of the injector that acts on always the same optical density $A_{N_1}(E)$.

\section{Enhancement of the spontaneous emission}
\label{enhancement}
As shown in previous sections, the spectra of our device are composed of a polaritonic contribution and a weak coupling luminescence. The scope of this section is to compare these two contributions and to discuss whether the electroluminescence can be enhanced in the strong coupling regime. To this aim, we take advantage of the angular separation between the two contributions.  In fact, as discussed in section~\ref{sample}, the polaritonic emission is mainly concentrated close to the fundamental photonic mode, hence in an angular interval between $66^{\circ}$ and $82^{\circ}$, while the weak coupling luminescence occurs mainly at smaller propagation angles.
Figure~\ref{integra}a shows, as a function of the internal angle, the voltage dependent ratio between the electroluminescence signal and the current, which is proportional to the quantum efficiency of the device. The most striking difference between these curves is the increase in the polaritonic quantum efficiency when decreasing the voltage applied to the structure: at 4 V the polaritonic quantum efficiency is twice the weak coupling one. A second difference between the curves is a small angular shift of the peak of the quantum efficiency ($+1.4^{\circ}$ when going from 4 V to 13 V). This is related to the shift of the injector energy with the voltage. These two aspects of the voltage evolution of the quantum efficiency are very well reproduced by our simulations, shown in fig.~\ref{integra}b, which is obtained by integrating at each angle the electroluminescence spectra of fig.~\ref{panel_sim}.
Note that the physical origin of the observed enhancement is a Purcell like effect, due to the increased photonic density of states of the fundamental mode with respect to the second order mode. The enhancement is hence obtained in our model by only considering the variation of the absorption spectrum from out of resonance to resonance angle (with a correction due to Fresnel coefficient). Indeed, the importance of the strong coupling regime is that it enables electroluminescent emission at energies which would be otherwise inaccessible, thanks to an electronic tunneling into the polariton states. This is well illustrated in figure~\ref{diff_coupes}, where the electroluminescence signal measured at 4.5 V is plotted for different constant energies: $E_{21}=161$~meV (blue triangles), 150 meV (red circles), 140 meV (black squares). These spectra are normalized to the weak coupling peak, in order to show that, thanks to the polariton dispersion, the electroluminescence signal is enhanced by a factor which depends on the energy position of the injector. An enhancement of a factor of four is obtained at 4.5 V for the energy 140 meV.

\section{Conclusions}
\label{concl}
In conclusion, we have presented a detailed study of the electroluminescence from an intersubband device operating in the light-matter strong coupling regime. We have compared two different devices with an identical active region, but fabricated with different optical resonators. This has allowed us to point out the peculiar features which are unique to the strong coupling regime. We have shown that the electroluminescence signal in the strong coupling regime is composed by two main contributions: the first originates from the lower polariton branch, while the second is due to the weak coupling of the intersubband transition with a second order cavity mode. Moreover, we have proven that the shape and the voltage evolution of the electroluminescence spectra are well reproduced with numerical simulations by considering that the polariton branches are populated only close to the energy of the injector. By studying the electroluminescence spectra at the energy of the intersubband transition, we have shown that the Rabi splitting does not vary with the voltage applied to the device. This implies that the population of the ground subband does not vary in the spanned voltage range. As a consequence, the voltage dependence of the electroluminescence spectra only results from the voltage dependence of the injector, i.e. from the position and width of the energy window in which the polariton states are populated. Notice that we did not observe any luminescence from the upper branch of the polariton dispersion. In fact above the alignment voltage (approximately 6~V) our band structure fixes the position of the injector in resonance with the state 2. Indeed, we have recently shown that the electroluminescence from the upper polariton can be observed by using a different design for the active region~\cite{delteil}. This work will be the object of a forthcoming publication.

The comparison between the weakly coupled and the polaritonic emission allowed us to show that the factor of two enhancement of the emission in the polaritonic mode is only due to a Purcell like effect, originating from the different quality factor of the fundamental and the excited photonic mode. Indeed, the importance of the strong coupling regime resides in the fact that it makes possible to achieve electroluminescence at energies otherwise inaccessible.
The quantum efficiency of polaritonic devices can be improved, in our opinion, with a more selective injection into the polariton states, far from the energy of the bare intersubband transition. This could be obtained by simultaneous engineering of the active region and the photonic dispersion, and by exploring different kind of resonators~\cite{todorov_PRL}.

\begin{acknowledgments}
The authors thank A. Delteil, C. Ciuti and M.~Za{\l}u\.{zny} for fruitful discussions. The device fabrication has been
performed at the nano-center \textit{Centrale Technologique
Minerve} at the Institut d'Electronique Fondamentale. This work has been partially supported by the French National Agency (ANR) in the frame of its Nanotechnology and Nanosystems program P2N, project ANR-09-NANO-007. We acknowledge financial support from the ERC grant ``ADEQUATE''.
\end{acknowledgments}

\newpage
\begin{figure}[ht]
\includegraphics[width=0.8\columnwidth]{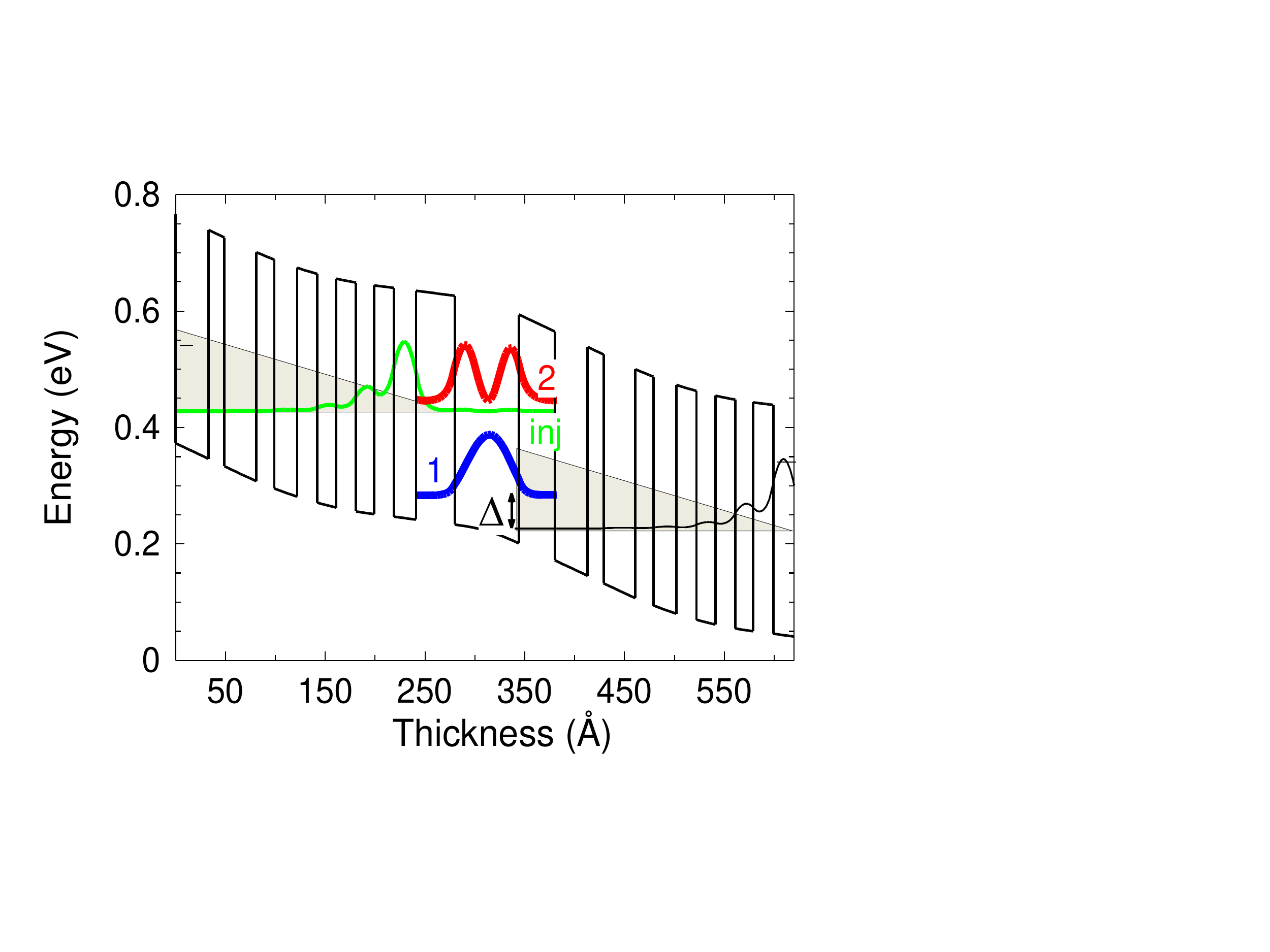}
\caption{(Color online) Band diagram of the quantum cascade structure at a voltage of 6~V obtained by solving coupled Schr\"{o}dinger and Poisson equations. The bold curves represent the fundamental (1) and excited (2) state of the radiative transition, as well as the injector state (inj). The energy difference between the state 1 and the injector quasi-Fermi energy is indicated by $\Delta$. The injection and extraction minibands are schematized by triangles. }
\label{bd}
\end{figure}

\begin{figure}[ht]
\includegraphics[width=0.8\columnwidth]{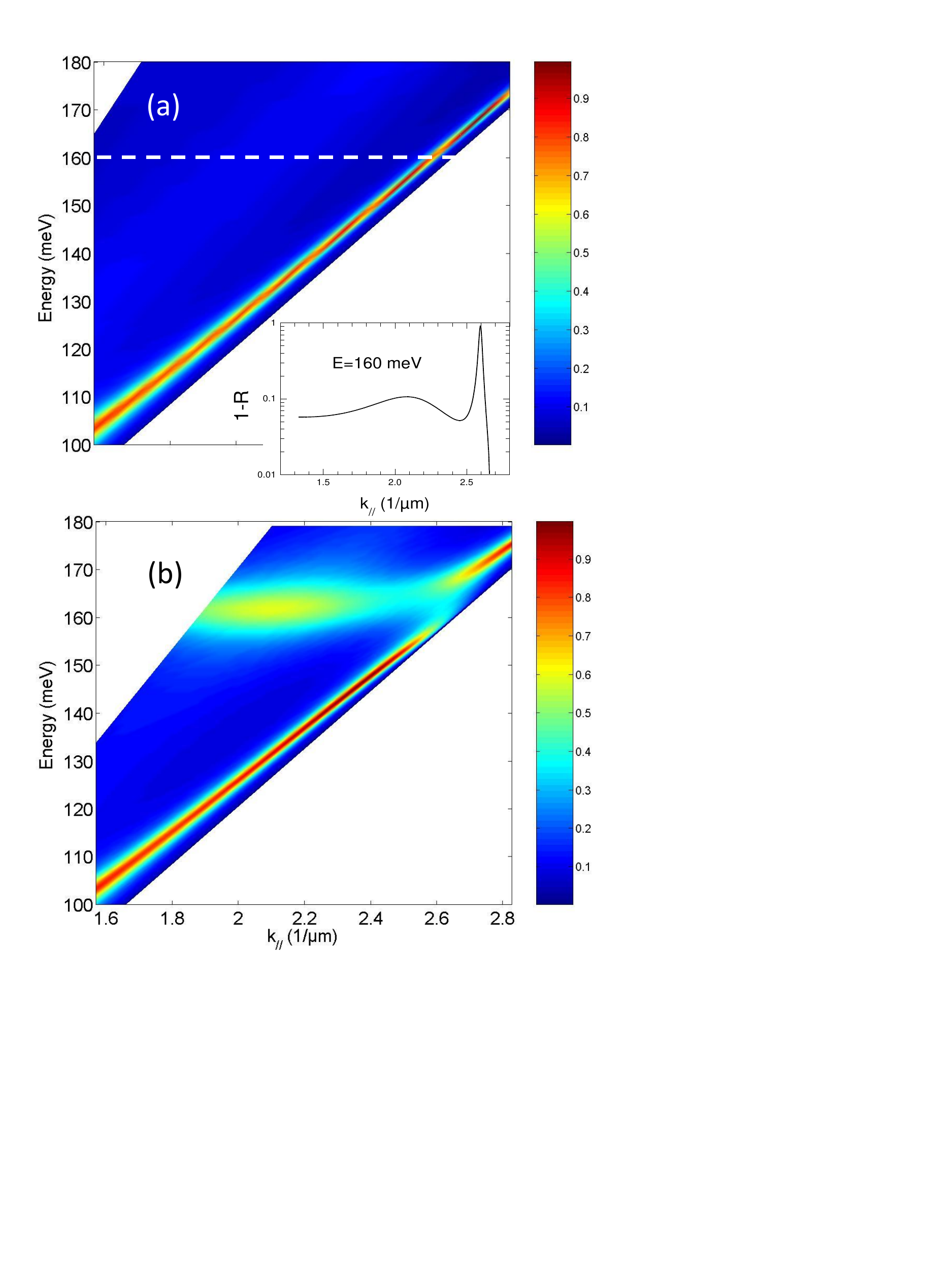}
\caption{(Color online) Absorption coefficient of the device with the top mirror: a) $1-R$ calculated as a function of the photon in-plane momentum and energy, without including the contribution of the intersubband transition. The horizontal dashed line indicates the energy used in the inset. b) Calculated absorption spectrum, including the contribution of the intersubband transition. The simulation has been obtained by using $E_{21}=161$ meV and $N_1=6 \times 10^{11} \rm{cm}^{-2}$. Inset: Simulated absorption (in logarithmic scale) at $E_p=160$~meV, plotted as a function of $k_{//}$.}
\label{absorption}
\end{figure}

\begin{figure}[ht]
\includegraphics[width=0.8\columnwidth]{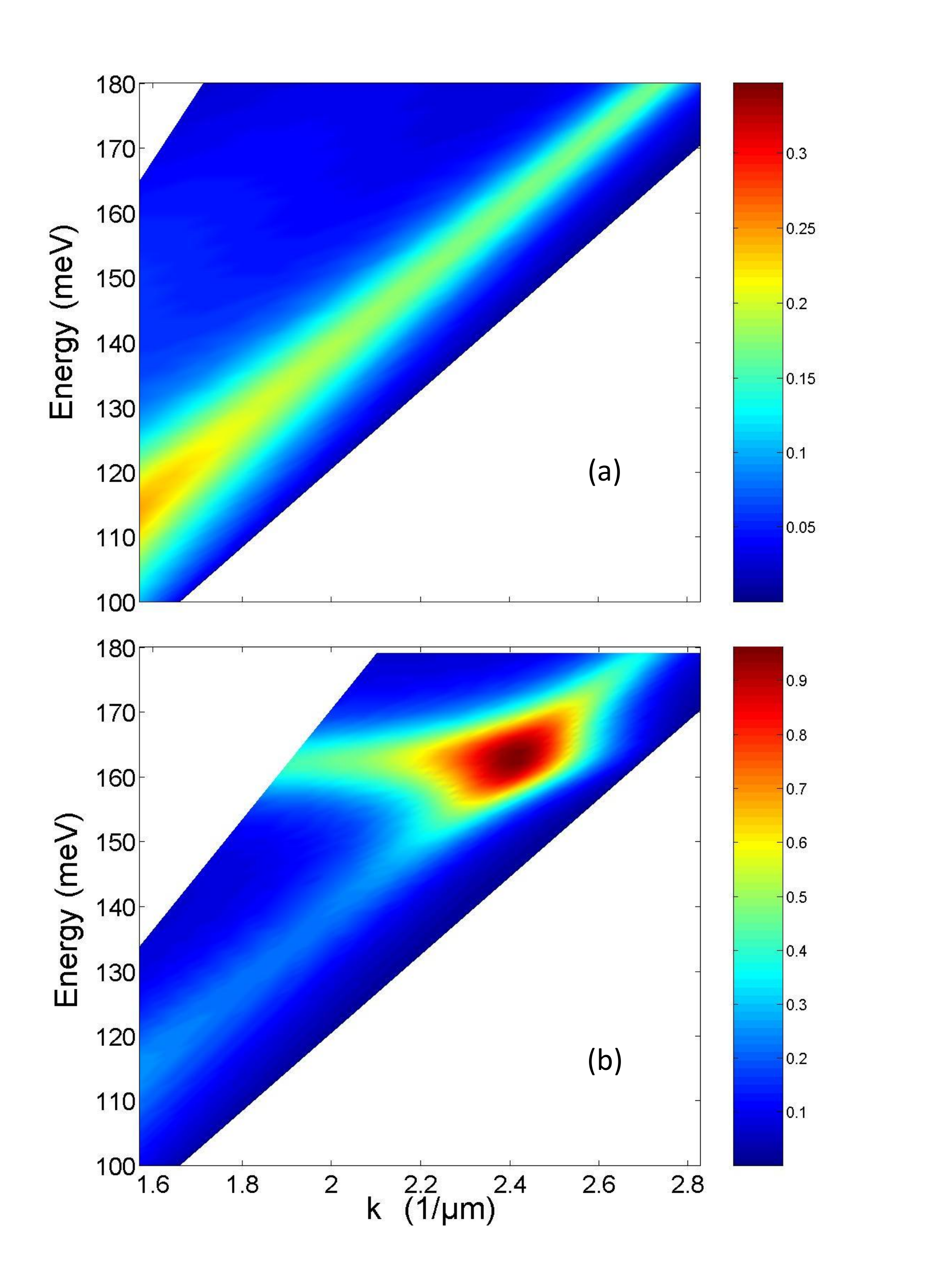}
\caption{(Color online) Absorption coefficient of the device without the top mirror (air on top): a) $1-R$ calculated as a function of the photon in-plane momentum and energy, without including the contribution of the intersubband transition. b) Calculated absorption spectrum, including the contribution of the intersubband transition.  }
\label{absorption_nomirror}
\end{figure}

\begin{figure}[ht]
\includegraphics[width=0.8\columnwidth]{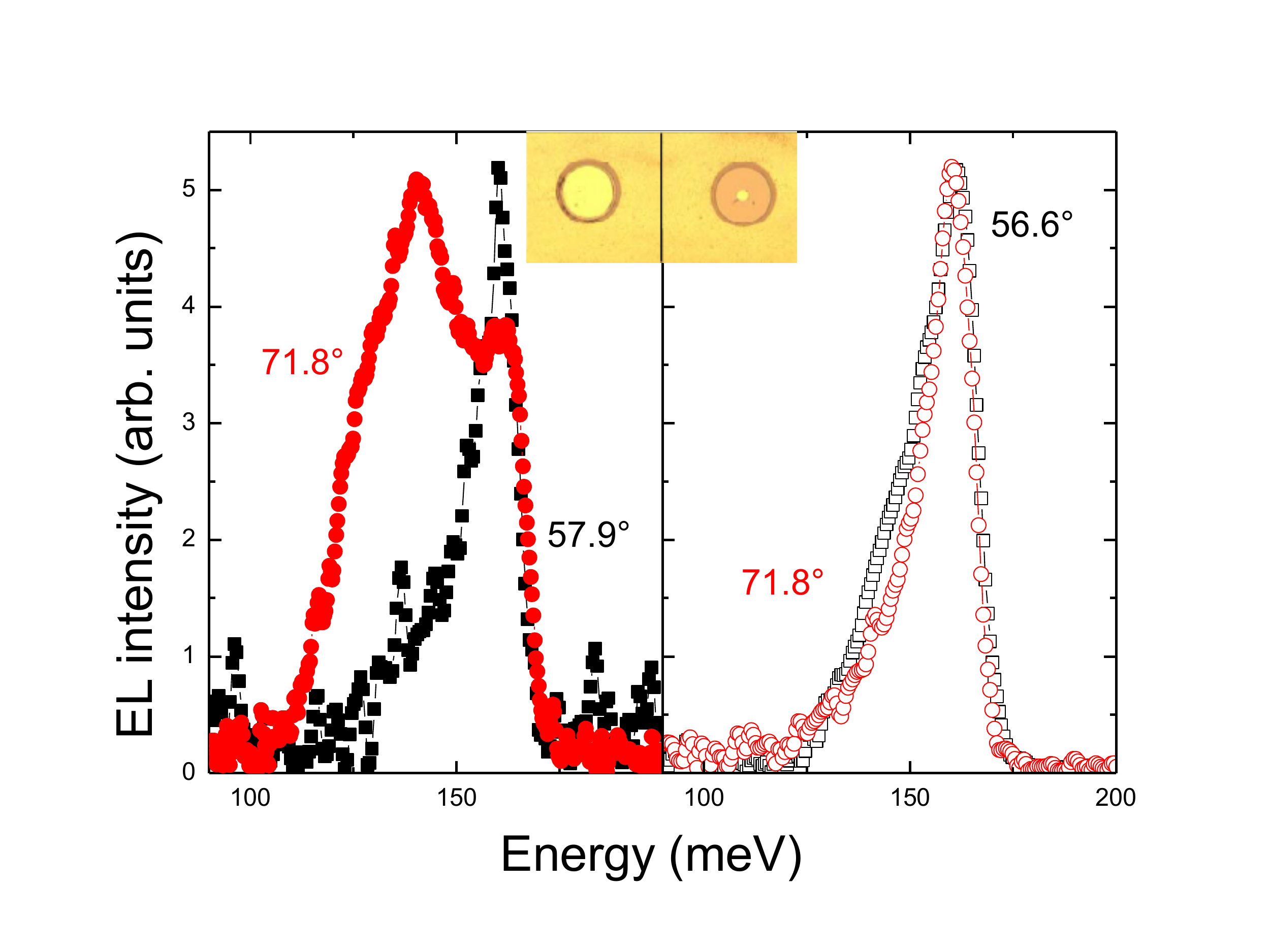}
\caption{(Color online) Electroluminescence spectra at 4.5 V (14 mA, 50$\%$ duty cycle) and 77 K for a device with (left panel) and without (right panel) top metallic contact. The curves indicated by squares have been obtained at a low value of the internal angle, while those indicated by circles have been obtained for an angle close to $\theta_{res}$. The values of the integrated optical power for the four spectra are the following: Left panel: 200 pW ($71.8^\circ$), 87 pW ($57.9^\circ$); Right panel: 68 pW ($71.8^\circ$), 120 pW ($56.6^\circ$). The inset is an optical microscope picture of the two mesa devices. }
\label{topEL}
\end{figure}

\begin{figure}[ht]
\includegraphics[width=0.8\columnwidth]{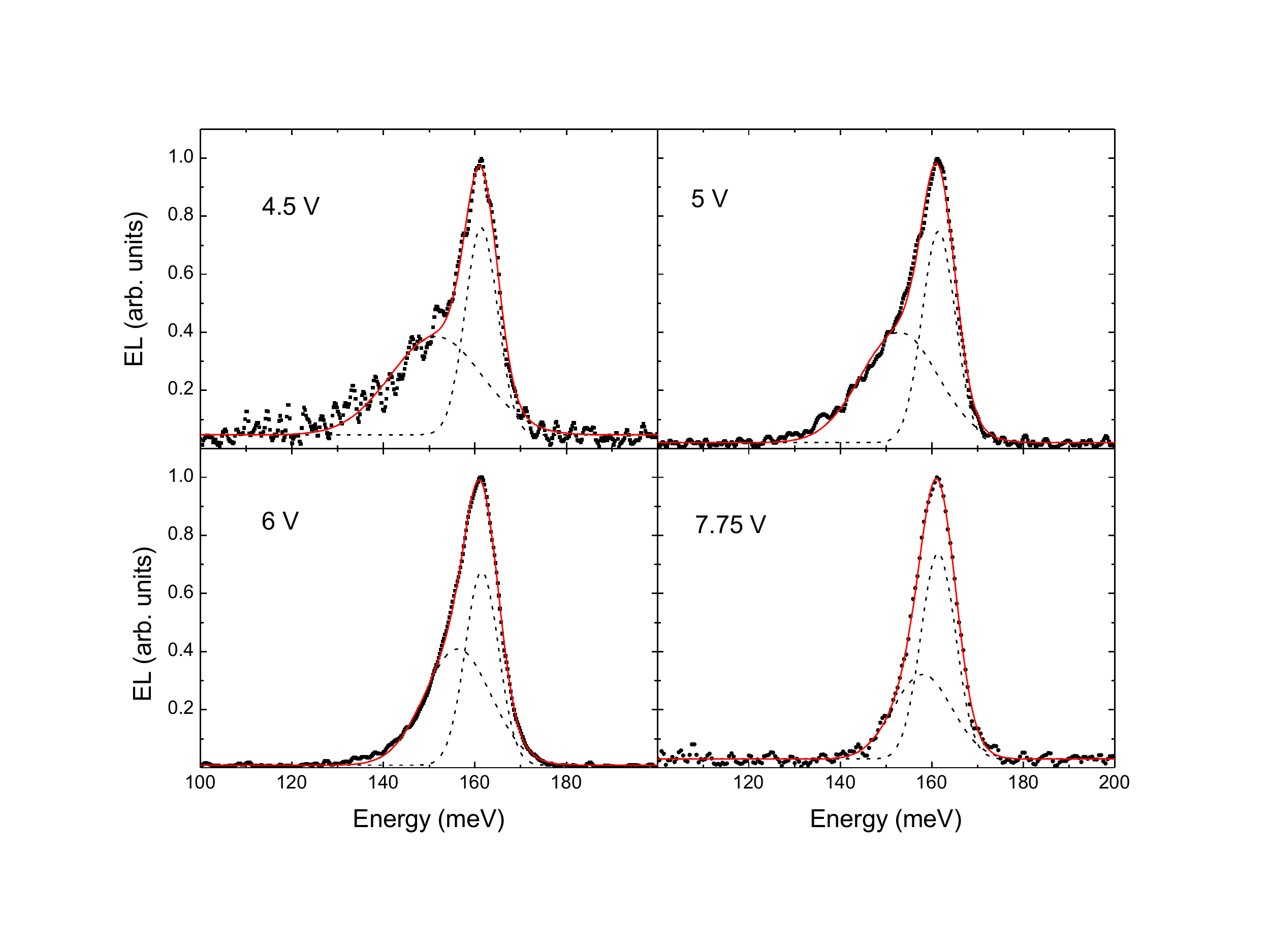}
\caption{(Color online) Electroluminescence spectra (symbol) measured at different voltages from the polaritonic device at an angle of approximately 58$^\circ$. The continuous lines are the best fit of the experimental data, given by the sum of two Gaussian functions, shown by dashed lines.}
\label{fits}
\end{figure}

\begin{figure}[ht]
\centering
\includegraphics[width=0.8\columnwidth]{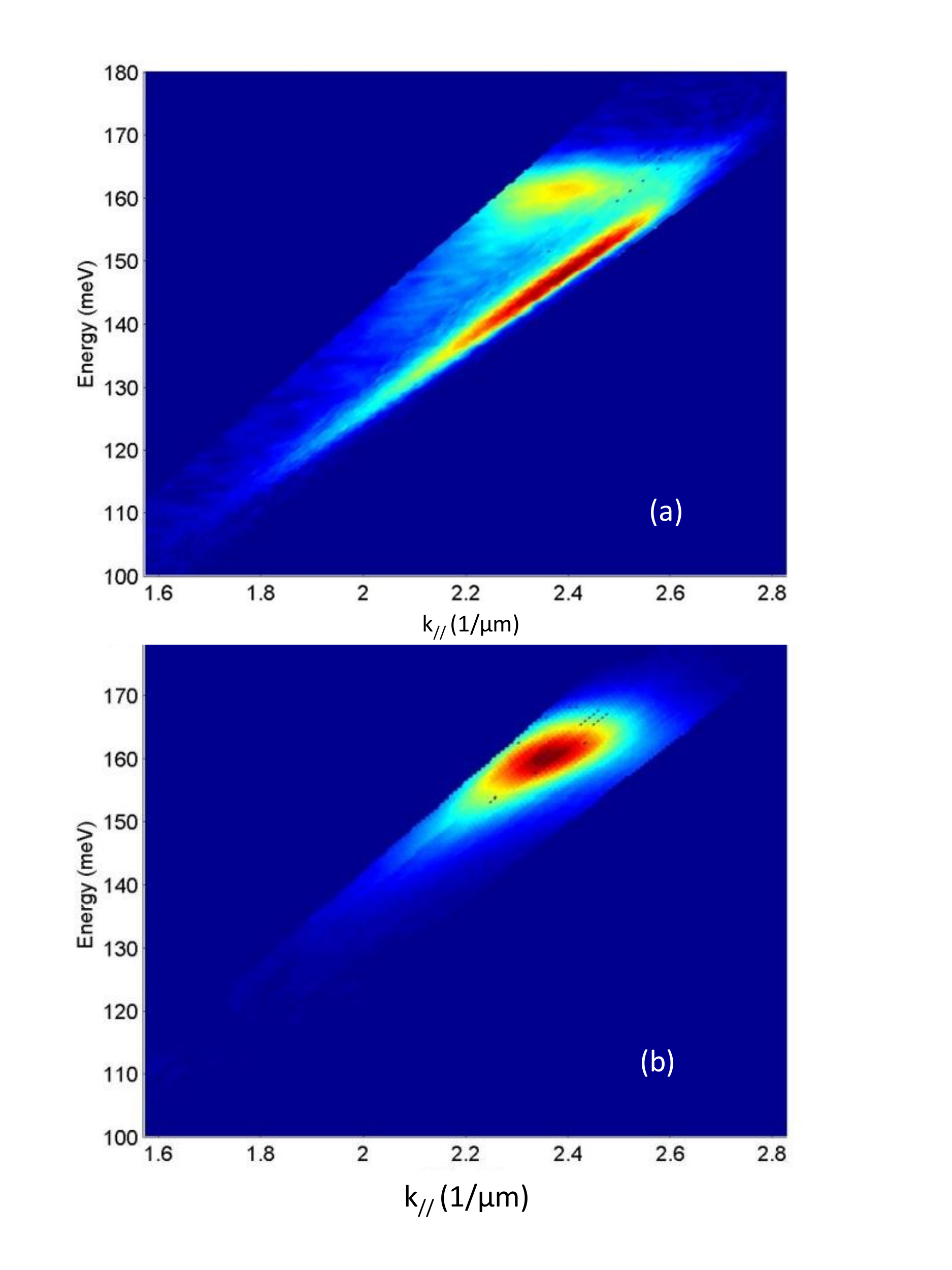}
\caption{(Color online) Electroluminescence as a function of the photon energy and in-plane momentum measured at 4.5 V (14 mA) and 77 K for the device with (a) and without (b) top metallic mirror. }
\label{EL_comp}
\end{figure}

\begin{figure}[ht]
\centering
\includegraphics[width=0.8\columnwidth]{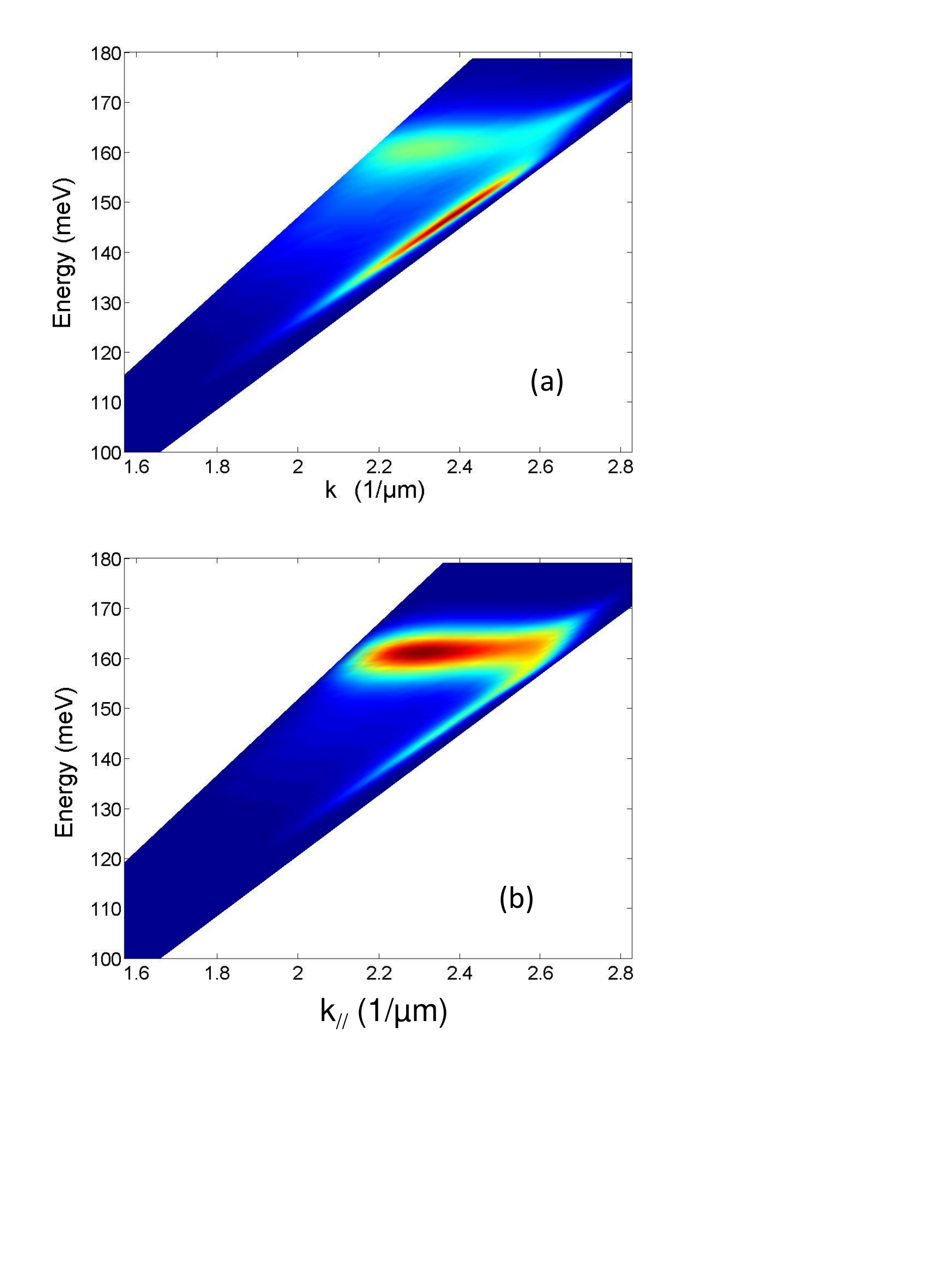}
\caption{(Color online) a) Contour plot of the simulated electroluminescence at 4.5 V, obtained by using eq.~\ref{sim_EL}. The injector energy position with respect to the fundamental subband is 150.5 meV and its width 12 meV.  b) "Photonic injection" simulated as the product of the electroluminescence spectrum in fig.~\ref{topEL}, left panel, at $52.4^{\circ}$ times the calculated absorption and the Fresnel coefficients. }
\label{twosimul}
\end{figure}

\begin{figure}[ht]
\includegraphics[width=0.8\columnwidth]{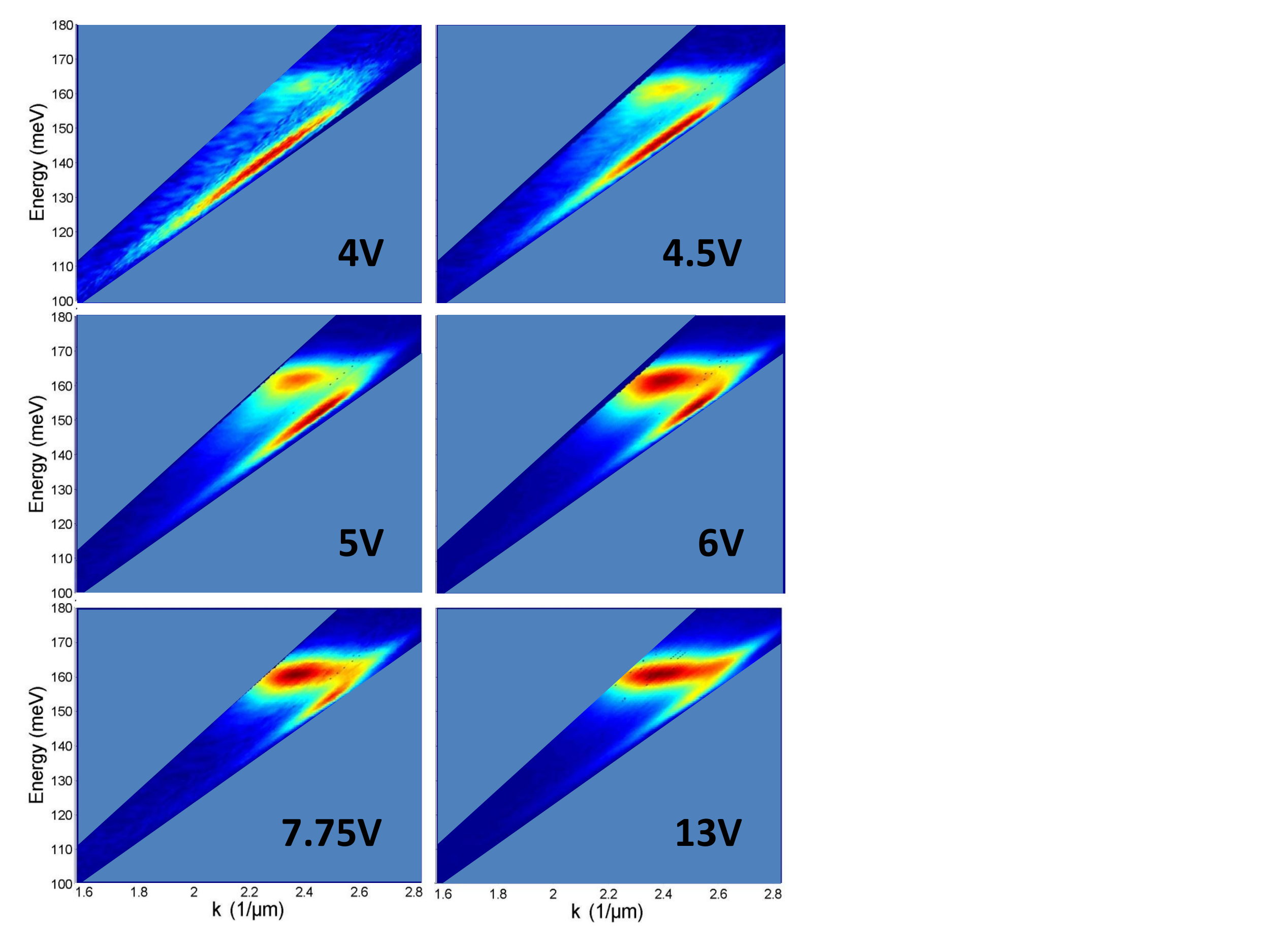}
\caption{(Color online) Contour plots of the electroluminescence spectra measured at different voltages. The values of the current for the voltages indicated in the figure are respectively: 4 mA, 14 mA, 58 mA, 258 mA, 718 mA, 2 A. The electroluminescence signal in the contour plots is normalized to one. The peak power is respectively (from the lowest to the highest voltage): 2.6 pW, 14 pW, 73 pW, 340 pW, 720 pW, 2.1 nW.}
\label{panel_meas}
\end{figure}

\begin{figure}[ht]
\includegraphics[width=0.8\columnwidth]{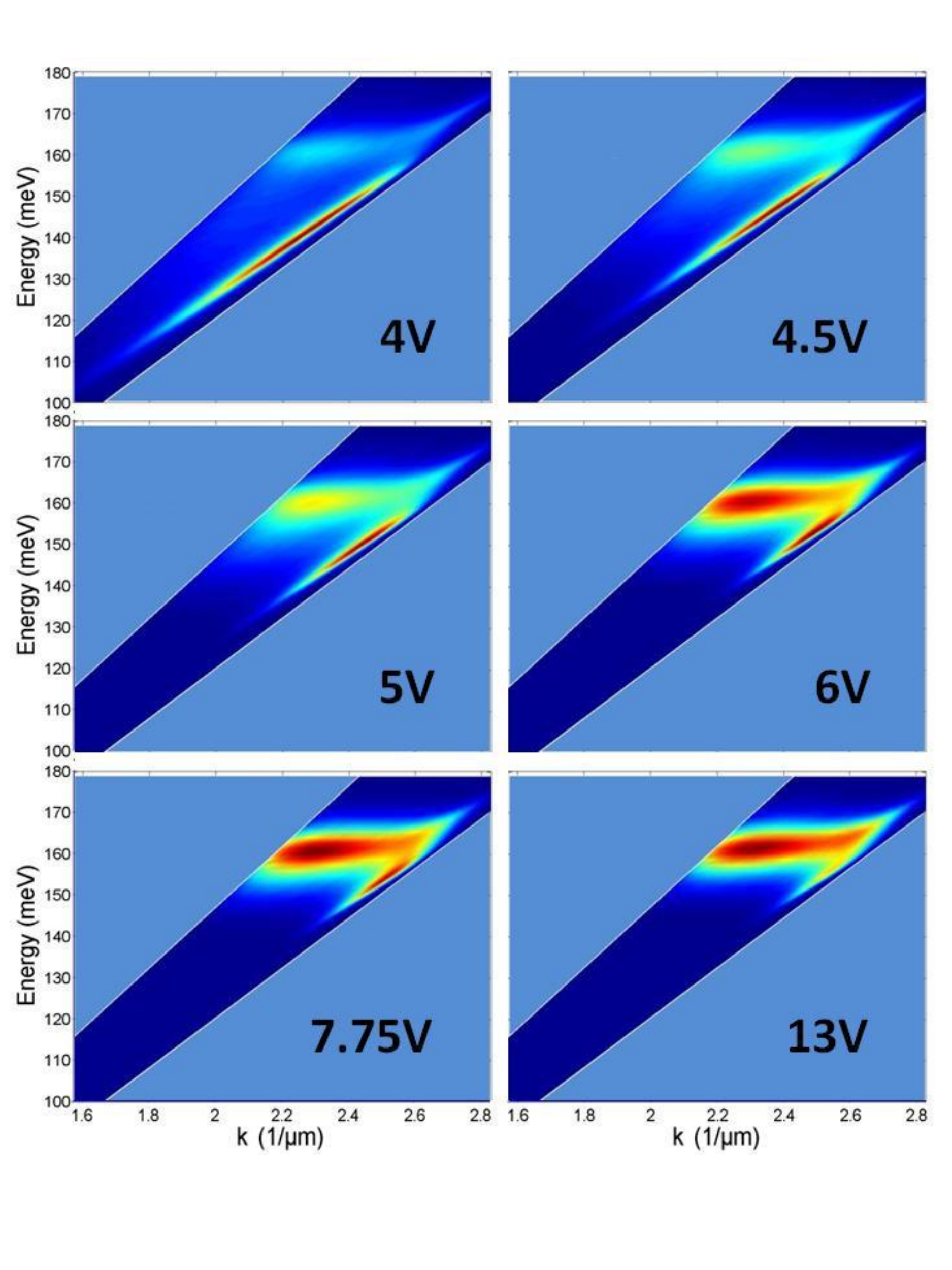}
\caption{(Color online) Contour plots of the simulated electroluminescence spectra at different voltages. The parameters used for the simulations are presented in table 1.}
\label{panel_sim}
\end{figure}

\begin{figure}[ht]
\includegraphics[width=0.95\columnwidth]{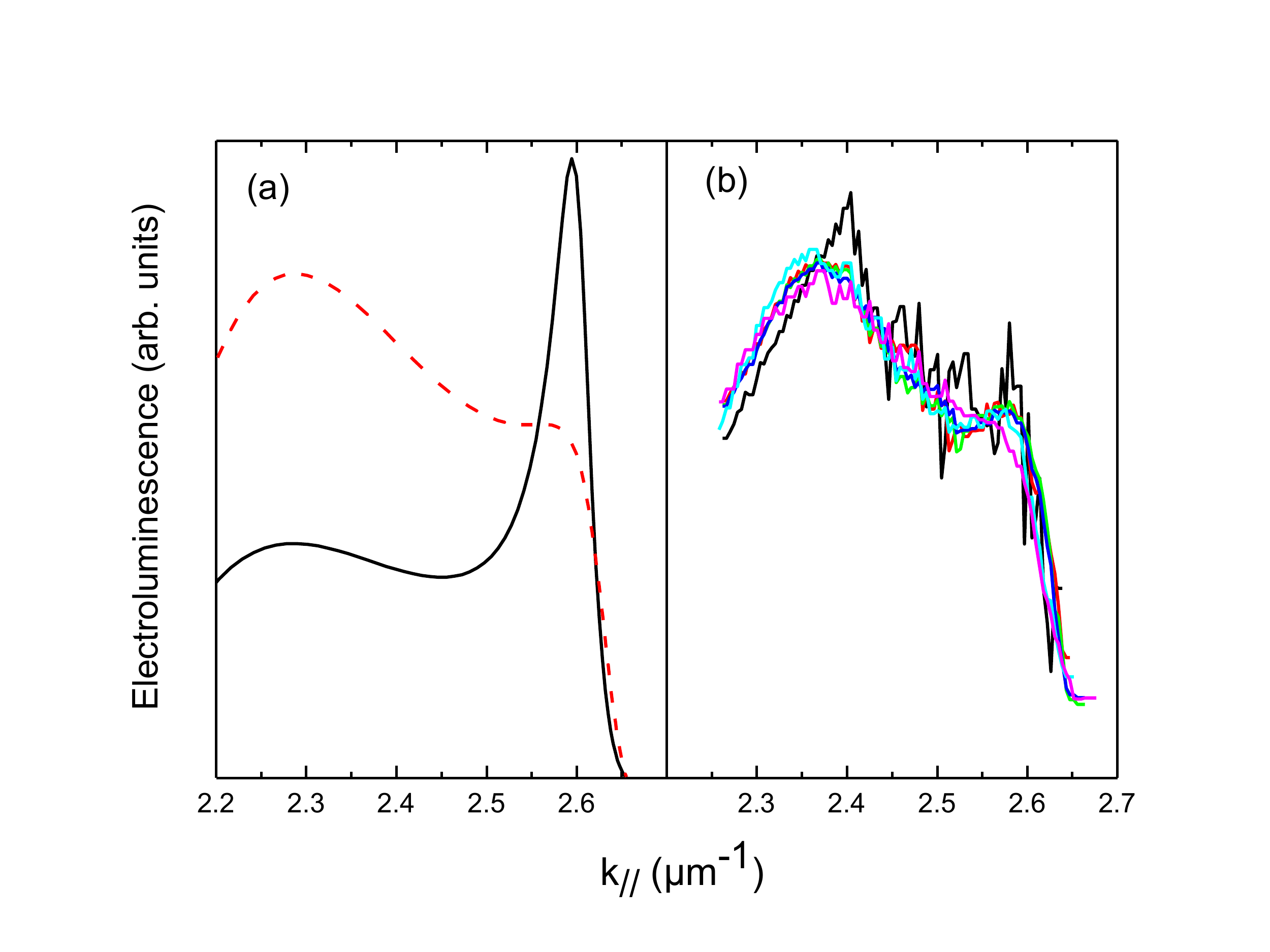}
\caption{(Color online) a) Simulated electroluminescence for the polaritonic device, obtained with $N_1=4 \times 10^{11} \, \rm{cm}^{-2}$ (red line) and $N_1=1 \times 10^{11} \, \rm{cm}^{-2}$ (black line). b) Normalized electroluminescence signal at the energy of the intersubband transition as function of the in-plane photon momentum, measured at different voltages: 4 V (black line), 4.5 V (red line), 5 V (blue line), 6 V (green line), 7.75 V (pink line), 13 V (light blue line).}
\label{coupes}
\end{figure}

\begin{figure}[ht]
\includegraphics[width=0.7\columnwidth]{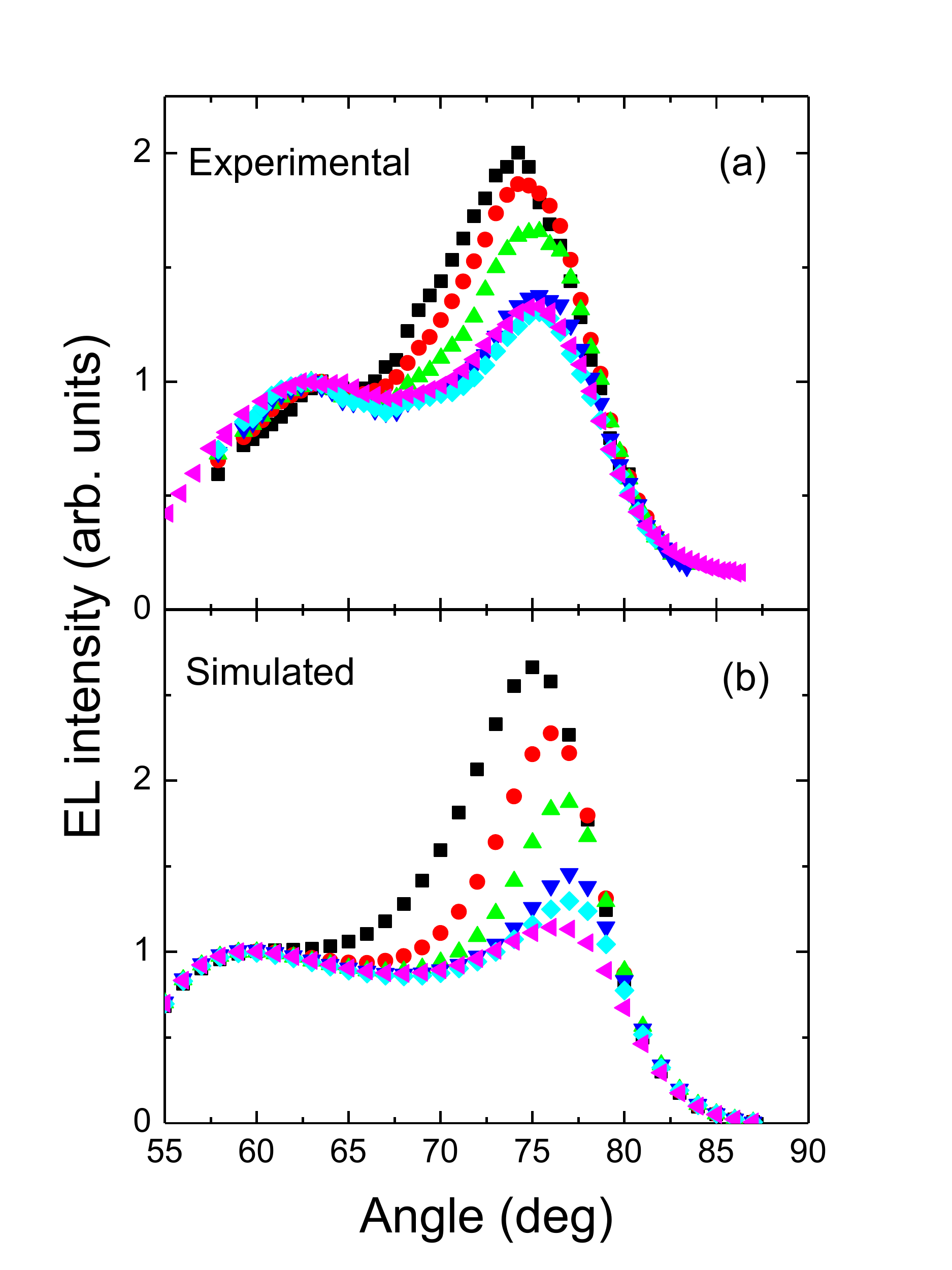}
\caption{(Color online) a) Integrated EL measured as a function of the internal angle of light propagation at different voltages: 4 V (black squares), 4.5 V (red circles), 5 V (green triangle), 6 V (blue down triangle), 7.75 V (light blue diamond), 13 V (pink left triangle). b) Simulation of the integrated EL at the same voltages as the experiments.}
\label{integra}
\end{figure}

\begin{figure}[ht]
\includegraphics[width=0.9\columnwidth]{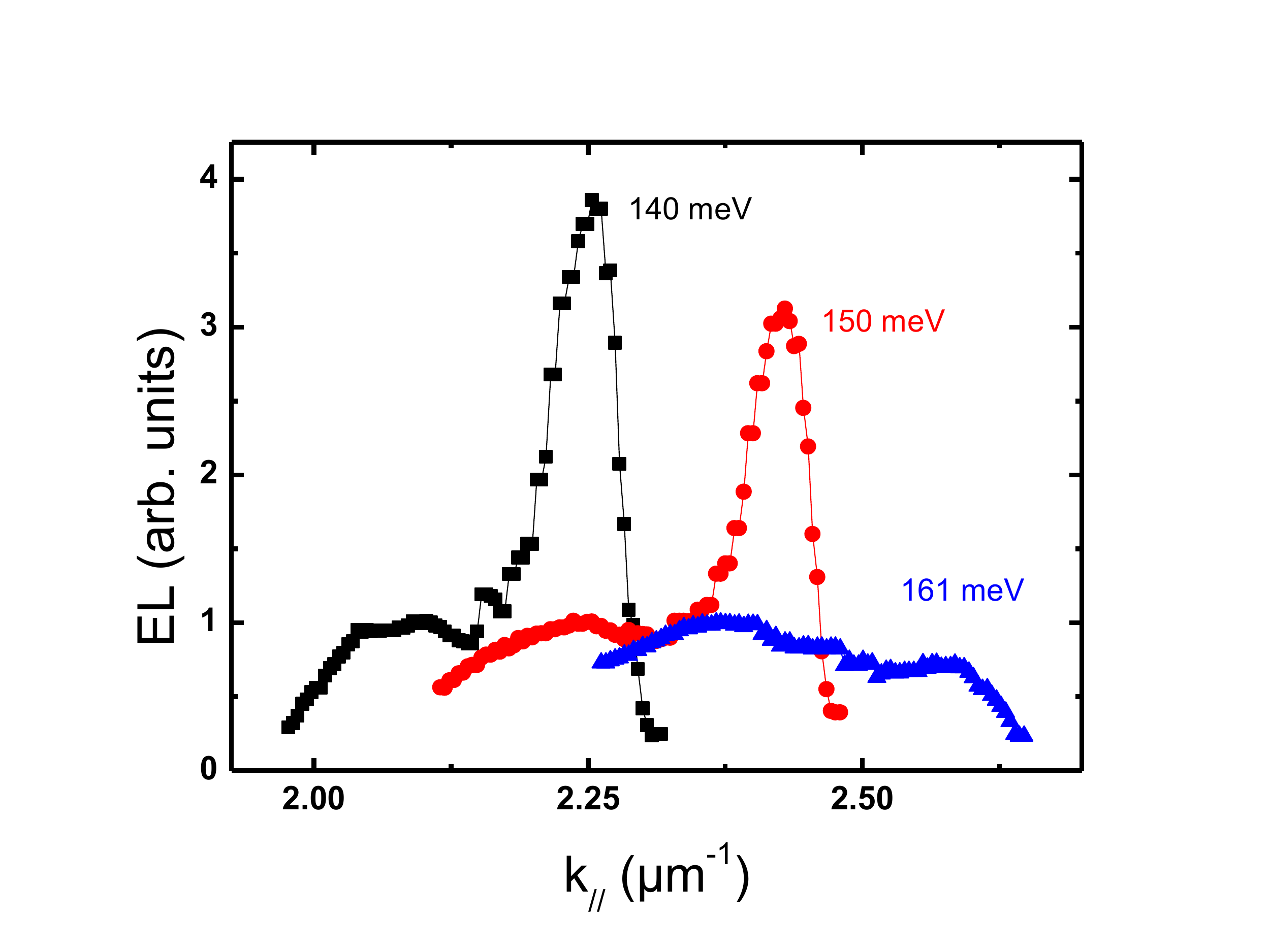}
\caption{(Color online) Electroluminescence spectra measured at 4.5 V at different energies: 161 meV (Blue triangles), 150 meV (red circles), 140 meV (black squares). The spectra have been normalized at the weak coupling peak, in order to evidence the polaritonic emission with respect to the weak coupling one.}
\label{diff_coupes}
\end{figure}


\begin{thebibliography}{32}
\bibitem{nota_fit} At 13~V, the electroluminescence spectrum is best fitted by a sum of a Lorentzian function centered at the energy $E_{21}$ and a Gaussian function for the diagonal transition.
\bibitem{nota} At a voltage of 4V, the electroluminescence measured at 58$^\circ$ is too weak to extract the energy position and width of the injector. The values that we used have been chosen to obtain the best agreement between the angle resolved simulated and measured electroluminescence spectra.
\end{thebibliography}
\end{document}